\documentclass[twoside]{ilcws07}
\usepackage[latin1]{inputenc}
\usepackage[dvips]{graphicx,epsfig,color}
\usepackage{wrapfig,rotating}
\usepackage{amssymb,amsmath,array}

\begin{document}
\title{
%%%%   Paper title goes here  %%%%%%%%%%%%%%
Monolithic Pixels R\&D at LBNL} %% 
%***********************************************************************
% AUTHORS INFORMATION AREA
%***********************************************************************
\author{Devis~Contarato$^1$, Marco~Battaglia$^{1,2}$, Jean-Marie~Bussat$^1$,\\
Peter Denes$^1$, Piero~Giubilato$^3$, Lindsay~Glesener$^2$,\\
Benjamin~Hooberman$^2$ and Chinh Qu Vu$^1$
% DO NOT MODIFY THE FOLLOWING '\vspace' ARGUMENT
\vspace{.3cm}\\
% Addresses and institutions (remove "1- " in case of a single institution)
1- Lawrence Berkeley National Laboratory \\
Berkeley, CA - USA
\vspace{.1cm}\\
2- University of California at Berkeley - Department of Physics \\
Berkeley, CA - USA\\
\vspace{.1cm}\\
3- Istituto Nazionale di Fisica Nucleare - Sezione di Padova \\
Padova - ITALY\\
}
%%***********************************************************************
% END OF AUTHORS INFORMATION AREA
%***********************************************************************
\maketitle

\begin{abstract}
This paper reports recent results from the ongoing R\&D on monolithic pixels
for the ILC Vertex Tracker at LBNL~\cite{url}. 
\end{abstract}

\section{Introduction}
A Laboratory Directed Research and Development (LDRD) program has been
on-going at LBNL since 2005. The main activity on sensor R\&D is the
development of CMOS monolithic pixels with integrated functionalities and
fast readout in view of their possible application in the ILC Vertex Tracker.
At each step of the development, various pixel architectures and layout
options are explored, driving the choice for the design of subsequent
prototypes. 

The results obtained with the first prototype designed at LBNL (the LDRD-1
chip) have been reported in~\cite{contarato_lcws05}. This paper summarizes
the recent results obtained with the second prototype of the LDRD family,
the LDRD-2 chip implementing in-pixel Correlated Double Sampling (CDS),
outlines the main features of a forthcoming prototype with integrated
digitization, and finally introduces a line of research recently started
on Silicon-On-Insulator (SOI) pixels.

\section{CMOS Monolithic Pixel Sensors}

\subsection{Prototype with in-pixel CDS and fast readout}
The LDRD-2 chip was designed and fabricated in 2006 using the AMS 0.35~$\mu$m
CMOS-OPTO technology with a nominal epilayer thickness of 14~$\mu$m.
The array of 96$\times$96 pixels of 20~$\mu$m pitch is divided in 6 subsections
with different sizes of the charge collecting diode (3$\times$3~$\mu$m$^2$
and 5$\times$5~$\mu$m$^2$) and different pixel architectures, standard
3-transistor (3T) pixels with and without a guard-ring around the diode and
a self-biased pixel architecture, similar to the one first proposed
in~\cite{strasbourg_selfbias}.

All sectors implement in-pixel CDS: the pixel dark level and charge signal
are consecutively stored on two capacitors integrated in each pixel; both
signals are then clocked to the chip analog outputs, after which the CDS
difference is obtained either via FPGA or via the online DAS. The chip was
designed to be operated up to a clocking frequency of 25~MHz. The readout
scheme is based on a rolling-shutter scheme, i.e. rows are reset and
read out in sequence, so that the integration time is constant for all pixels.

The LDRD-2 has been extensively tested using the same measurement
protocol used for the LDRD-1 chip: in-lab calibrations with a $^{55}$Fe
source and lasers of different wavelengths have been performed. The sensor
charge collection time has been determined using a fast (0.5-5~ns)
1060~nm laser pulse focused on a single pixel and by following the collected
charge signal during the pixel sampling time. The result of $\sim$150~ns
agrees well with the expectations coming from the typical resistivity of the
epitaxial layer employed in the fabrication process.

\begin{figure}[t]
\centerline{\includegraphics[width=\columnwidth]{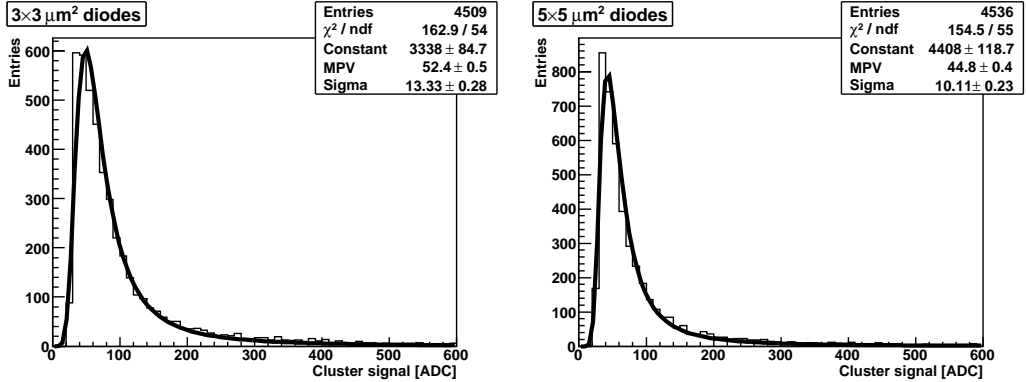}}
\caption{LDRD-2 chip beam-test results: cluster signal for 1.5~GeV electrons for
two different sizes of the charge collecting diode.}
\label{fig:ldrd2_als}
\end{figure}

The sensor response to minimum ionising particles has been tested on the
1.5~GeV electron beam at the BTS line of the Advanced Light Source (ALS)
at LBNL. Figure~\ref{fig:ldrd2_als} shows the cluster signal distribution
obtained for the two different diode sizes on the chip sectors featuring a
3T pixel architecture. An average S/N ratio of $\sim$20 was found to be
similar for the two different diode sizes, with an average cluster multiplicity
between 4 and 5, the larger being associated with pixels equipped with smaller
diodes. The most probable value for the cluster charge signal for small and big
diodes is $\sim$750 and $\sim$1000 electrons, respectively.

\subsection{Prototype with integrated ADCs}
The LDRD-3 chip, which is being designed and will be submitted in July
in the 0.35~$\mu$m AMS CMOS-OPTO process, features the same in-pixel CDS
architecture as in the LDRD-2 prototype, combined with integrated ADCs
at the end of each column. The pedestal and signal levels from each pixel
are sent to the bottom of the column, the CDS difference being performed
by a 5-bit successive approximation, fully differential ADC running at
300~MHz clock frequency. The design of the LDRD-3 chip is driven by the
results of a detailed study performed on data collected with the LDRD-1
chip for different pixel pitches, which had shown that 5-bit ADC accuracy
is sufficient to achieve the desired single point resolution, provided
that the pixel pedestal can be removed before digitization.

\begin{figure}[t]
\begin{center}
\begin{minipage}{.49\linewidth}
\centerline{\includegraphics[width=\linewidth]{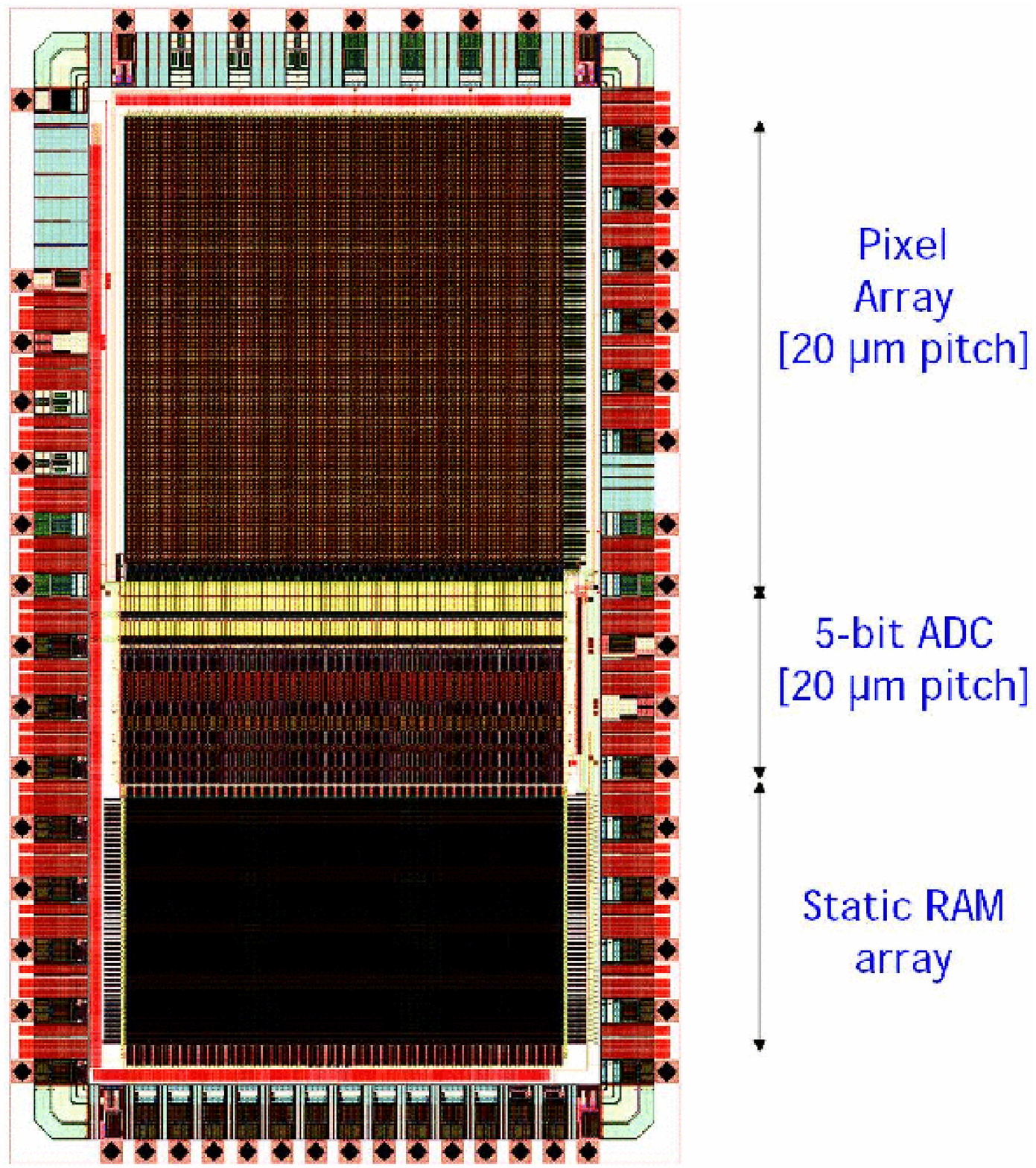}}
\end{minipage}\hfill
\begin{minipage}{.49\linewidth}
\centerline{\includegraphics[width=\linewidth]{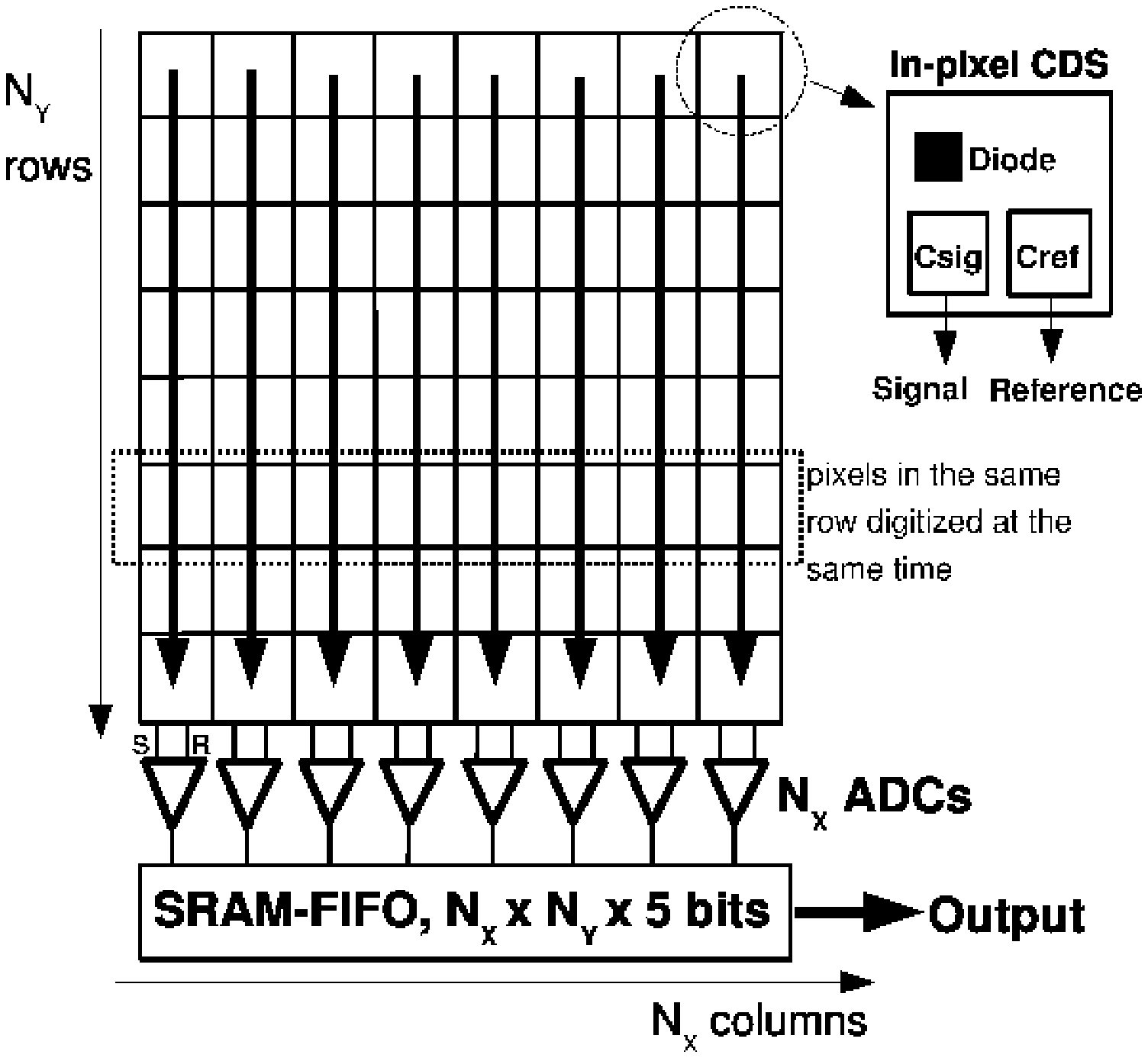}}
\end{minipage}
\caption{Layout (left) and schematic (right) of the LDRD-3 chip.}
\label{fig:ldrd3}
\end{center}
\end{figure}

Figure~\ref{fig:ldrd3} shows the chip layout and a sketch illustrating
the chip working principle. The array consists of 96$\times$96 pixel of
20~$\mu$m pitch. After a global reset, the sensor image is integrated,
and then the pixel signal and reference levels are stored in the in-pixel
capacitors. This is followed by a column-wise digitization of the pixel
signals with concurrent storage in a SRAM-FIFO memory cell integrated at
the bottom of the column. The digital image is then read out serially
from the SRAM-FIFO using a 50~MHz clock. The chip is expected to be back
from fabrication in October 2007.

\section{SOI monolithic pixels}
A first prototype chip has been designed and submitted in a 0.15~$\mu$m
fully-depleted SOI technology from OKI~\cite{oki} in Fall 2006. The sensor
is based on a 350~$\mu$m thick high-resistivity substrate, separated by the
CMOS circuitry by a 200~nm thick buried oxide. The CMOS circuitry is
implanted on a 40~nm layer of silicon on top of the buried oxide. The
thickness of this CMOS layer is small enough that the layer can be fully
depleted. This technology is expected to combine the advantage of a depleted
substrate with the possibility of integrating complex functionalities by
making use of both types of MOS transistor, in contrast with standard
CMOS monolithic APS which feature only $n$-type transistors. Moreover,
higher speed and lower power consumption can be achieved with the SOI process
with respect to conventional CMOS processes.

The chip features 10~$\mu$m pitch pixels, two analog sections with 1.0~V and
1.8~V bias and a digital section; all three sections are divided in 2
subsections with 1$\times$1 and 5$\times$5~$\mu$m$^2$ charge collecting
diodes. The total array size is of 160$\times$150 pixels. The chip has been
received back from foundry in May 2007. It is fully functional and is
currently being tested. 

\section{Conclusions \& Outlook}
Recent progress in the development of the LDRD monolithic pixel chips at
LBNL has been reviewed. A prototype with in-pixel CDS and fast readout
has been successfully tested and will be used in June-July for tracking
studies with 120~GeV protons at the MTBF facility at FNAL, in conjuction
with the Thin Pixel Telescope described in~\cite{battaglia}. Irradiation
tests with 30~MeV protons and 1-20~MeV neutrons are also planned in Summer.
The next prototype, implementing on-chip digitization, is expected in Fall
2007. In parallel, tests have started on a novel SOI chip, which combines
a high-resistivity sensor substrate with a full CMOS circuitry implanted
on top, so that higher circuitry complexity with low power dissipation can
be implemented. 

\section*{Acknowledgments}
This work was supported by the Director, Office of Science, of the US
Department of Energy under Contract no. DE-AC02-05CH11231.

\begin{footnotesize}

\end{footnotesize}


\begin{thebibliography}{99}

\bibitem{url} Slides: \\ 
\verb$http://ilcagenda.linearcollider.org/contributionDisplay.py?contribId=302&sessionId=74&confId=1296$

\bibitem{contarato_lcws05} D.~Contarato {\it et~al.}, Proceedings of the LCWS06 International Linear Collider Workshop 2006, Bangalore, India.
\bibitem{strasbourg_selfbias} G.~Deptuch {\it et~al.}, Nucl. Instrum. Meth. {\bf A 511} (2003) 240. 
\bibitem{oki} OKI Electric Industry Co., Ltd., \verb$http://www.oki.com$
\bibitem{battaglia} M.~Battaglia {\it et~al.}, \emph{Particle Tracking with a Thin Pixel Telescope}, these Proceedings.


\end{thebibliography}
\end{document}